\documentclass[fleqn,usenatbib]{mnras}

\usepackage{newtxtext,newtxmath}

\usepackage[T1]{fontenc}
\usepackage{ae,aecompl}

\usepackage{pgfplotstable}
\usepackage{csvsimple}
\usepackage{booktabs} 

\usepackage{soul}

\usepackage{graphicx}	
\usepackage{amsmath}	

\newcommand{\Msun}{{$M_{\odot}$}}

\title[MIR imaging of SN\,1987A]{Mid-infrared imaging of Supernova\,1987A}

\author[M. Matsuura et al.]{
Mikako Matsuura,$^{1}$ \thanks{E-mail: matsuuram@cardiff.ac.uk (MM)}
Roger Wesson,$^{2, 1}$ 
Richard G. Arendt,$^{3, 4}$ 
Eli Dwek,$^{3}$ 
            \newauthor 
James M. De Buizer,$^{5}$
John Danziger,$^{6}$  
Patrice Bouchet,$^{7,8}$ 
M.J. Barlow,$^{2}$ 
Phil Cigan,$^{9}$
            \newauthor 
Haley L. Gomez,$^{1}$
Jeonghee Rho,$^{10}$ 
Margaret Meixner,$^{11, 12}$ 
\\
$^{1}$ School of Physics and Astronomy, Cardiff University, Queen's Buildings, The Parade, Cardiff, CF24 3AA, UK \\
$^{2}$ Department of Physics and Astronomy, University College London (UCL), Gower Street, London WC1E 6BT, UK \\
$^{3}$ Observational Cosmology Lab, Code 665, NASA Goddard Space Flight Center, Greenbelt, MD 20771, USA \\
$^{4}$ University of Maryland-Baltimore County, Baltimore, MD 21250, USA \\
$^{5}$ SOFIA-USRA, NASA Ames Research Center, MS 232-12 Moffett Field, CA 94035, USA\\
$^{6}$ Osservatorio Astronomico di Trieste, Via Tiepolo 11, Trieste, Italy \\
$^{7}$ DRF/IRFU/DAp, CEA-Saclay, F-91191 Gif-sur-Yvette, France \\
$^{8}$ CNRS/AIM, Paris-Saclay University, France \\
$^{9}$ Department of Physics and Astronomy, George Mason University, 4400 University Dr
Fairfax, VA 22030-4444, USA \\
$^{10}$ SETI Institute, 189 N. Bernardo Avenue, Mountain View, CA 94043, USA \\
$^{11}$ Space Telescope Science Institute, 3700 San Martin Drive, Baltimore, MD 21218, USA \\
$^{12}$ Department of Physics and Astronomy, The Johns Hopkins University, 366 Bloomberg Center, \\
3400 N. Charles Street, Baltimore, MD 21218, USA \\
}

\date{Accepted XXX. Received YYY; in original form ZZZ}

\pubyear{2022}

\hypersetup{draft}

\begin{document}
\label{firstpage}
\pagerange{\pageref{firstpage}--\pageref{lastpage}}
\maketitle

\begin{abstract}
At a distance of 50\,kpc, Supernova 1987A is an ideal target to study how a young supernova (SN) evolves in time.
Its  equatorial ring, filled with material expelled from the progenitor star about 20,000 years ago,  has been engulfed with  SN blast waves.
Shocks heat dust grains in the ring, emitting their energy at mid-infrared (IR) wavelengths
We present ground-based  10--18\,$\mu$m monitoring of the ring of SN\,1987A from day 6067 to 12814 at a resolution of 0.5'', together with {\it SOFIA} photometry at 10--30\,$\mu$m. 
The IR images in the 2000's (day 6067--7242) showed that the shocks first began brightening the east side of the ring.
Later, our mid-IR images from 2017 to 2022  (day 10952--12714) show that dust emission is now fading in the east,
while it has  brightened on the west side of the ring.
Because dust grains are heated in the shocked plasma, which can emit X-rays,  the IR and X-ray brightness ratio represent shock diagnostics.
Until 2007 the IR to X-ray brightness ratio remained constant over time, and during this time shocks seemed to be largely influencing the east side of the ring.
However, since then, the IR to X-ray ratio has been declining, due to increased X-ray brightness. 
 Whether the declining IR brightness is because of dust grains being destroyed or being cooled in the post-shock regions will require more detailed modelling.
\end{abstract}



\section{Introduction}

Supernovae (SNe) play a dual role in the evolution of interstellar dust. 
On one hand, they are the most important source of dust production in galaxies, but on the other had had also the most important source of grain destruction. 
Theoretical models show that most of the heavy elements produced can precipitate out of the gas and form refractory grains \citep{Sarangi:2013bj, Sarangi:2015he, 2019supe.book..313S, Sluder.2018ujo}. 
Infrared and submilimetre observations of Cassiopeia\, A \citep{Arendt:2014ka, Barlow:2010p29287, DeLooze:2017cz}, SN\,1987A \citep{Matsuura:2011ij, Indebetouw:2014bt, Matsuura:2015kn},  Crab Nebula \citep{Gomez:2012fm}, and young Galactic  (up to $\sim$2000 years old) SN remnants \citep{Chawner:2019dn} confirm the presence of $\sim$0.1--1.0\,\Msun\,  of dust,  indicating that a substantial fraction of refractory elements in their ejecta went to dust grains.
If the majority of dust in SNe can survive the shock interactions, SNe could be an important source of dust production in the ISM \citep{Dwek:2011p29471}.
%
The fate of this newly-formed dust is still a subject of active studies.
The reverse shock traveling through the ejecta can destroy newly-formed dust
\citep{Dwek.19964y, schneider04, Nozawa:2007kh, Biscaro:2014kh, Biscaro.2016, Silvia:2010p29877, Micelotta:2016jk, Kirchschlager.2019yrq}.
Any grains surviving the reverse shock may also be destroyed during the injection phase into the interstellar medium  \citep[ISM; ][]{Slavin.2020}
Thereafter, ISM dust will be subject to destruction as it encounters the SN remnant shocks. 
The grain destruction efficiency and ISM dust lifetimes are highly uncertain since they depend on a long list of parameters. 
Macroscopic parameters include, the energy of the SN explosion, the morphology of the medium surrounding the SN \citep{Slavin.2020} and that of the general ISM.  
Microscopic parameters include the composition and size distribution of the SN condensates, and the detailed interaction of the dust with the shocked gas and other grains 
\citep{Dwek:1992da, Jones:1996p8422, Slavin:2015in, Kirchschlager.2021, Priestley.2021}.
Because the evolution of dust in the ISM is a fine balance between dust production and destruction, intense investigations are currently underway into dust production and destruction by SNe. 
In this paper, we examine the latter point of view, and investigate how dust grains are impacted by SN shocks over time.

\begin{figure*}
	\includegraphics[width=18cm]{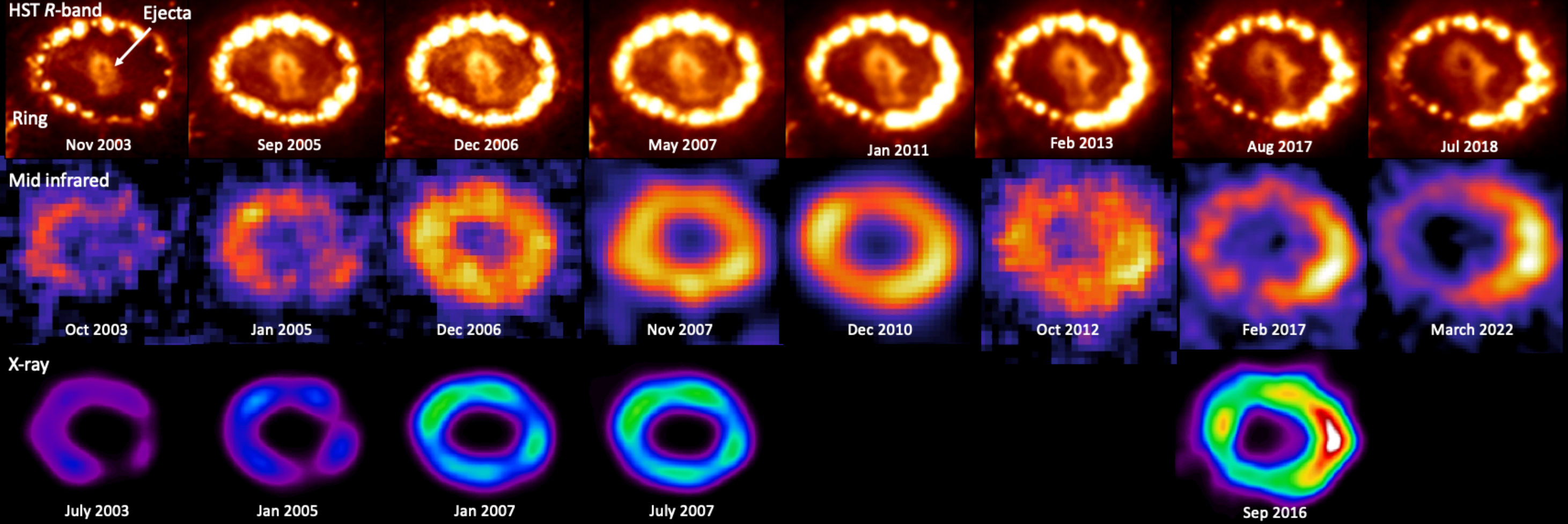}
    \caption{Time evolution of SN 1987A images, showing how the blast waves progress in time. The north is top and the east is left.
    The north-east side of the ring brightened, in early 2003, and  its brightness increased in 2000's, as the blast waves gradually passed with the brightest blobs at the north-east side of the ring. 
    In meanwhile, in 2010's images, the west side of the ring is the brightest.
    {\it HST} $R$-band images are from \citet{Larsson2019}, mid-infrared images from \citet{Bouchet:2006p2168} and this work, and X-ray 0.3--8\,keV images from \citet{Frank:2016ka} and Frank \& Burrows (private communication). 
   Note that mid-infrared images were taken with different filters around 10\,$\mu$m and have large uncertainties in fluxes, hence, the relative brightness at different epochs is  not accurate. 
            \label{fig:images}}
\end{figure*}

At a distance of  only 50\,kpc, SN 1987A provides a unique opportunity  for many discoveries, and has been monitored at almost all wavelengths as the SN remnant (SNR) has evolved over the past 30 years \citep{McCray:1993p29839, McCray:2016bg}.
The {\it Hubble Space Telescope (HST)} optical images  showed that the SN remnant is composed of ejecta and the circumstellar rings \citep[Fig.\,\ref{fig:images}; e.g.][]{Fransson:2015gp, Larsson:2016bj, Kangas.2021}. 
The circumstellar rings,  which are made of the equatorial ring and the outer rings, consist of the material lost from the progenitor  SK $-$69 202, when the star was in the red-supergiant phase about 20,000 years ago \citep{McCray:1993p29839}. 
Eventually,  the merger of a $\sim$15\,\Msun\, and a $\sim$5\,\Msun\, binary \citep{Morris.2007} created a blue-supergiant, and 
the circumstellar ring had been polluted  by the material from faster but thinner  wind from the blue supergiant.
The outer ejecta, which were expelled by the SN explosion and are mainly composed of hydrogen and helium, expand at 2,000--10,000 km\,s$^{-1}$
\citep{Larsson:2016bj, Kangas.2021}.
The  equatorial ring expands much slowly  \cite[$\sim$100\,km\,s$^{-1}$; ][]{Larsson2019}, and has been caught up by the outer ejecta, experiencing shocks.

The equatorial  ring,  hereafter the ring, in SN\,1987A provides an excellent place to study dust grains undergoing shocks.
Combined X-ray and infrared (IR) observations can provide important information of the interaction between dust grains and the shocked X-ray emitting gas \citep{Dwek:2010kv}. 
Mid-IR observations of SN\,1987A 
have shown the presence of hot, $\sim$180 K, silicate dust \citep{Bouchet:2004bs, Bouchet:2006p2168, Dwek:1987p3350, Dwek:2008p28793}, presumably collisionally heated by the shocked gas. 
The plasma that heats the dust and gives rise to the X-ray emission also destroys the dust that is swept up by the SN shock. 

In this paper, we present mid-IR monitoring  from day 6067--12814 of the ring images at a resolution of $\sim$0.5''.
These details allow us to follow the simultaneous evolution of the X-ray and IR emission of the different sections of the ring over time. 
These images and fluxes are compared with collisionally heated dust models, whose global parameters have been constrained by X-ray measurements of the plasma conditions.

\section{Observations}

\subsection{VLT/VISIR observations in 2017, 2018 and 2022}

The VLT Spectrometer and Imager for the mid-InfraRed \citep[VISIR; ][]{2004Msngr.117...12L, 2015Msngr.159...15K}  obtained images of SN 1987A, under the program 298.D-5023(A).  
The images of SN 1987A were taken  around the 30 year anniversary mark on 2017 Feb 15th (day 10,950 since the explosion), Feb 17th (day 10,952) and March 13th (day 10,976).
All were obtained with the B10.7 (10.7\,$\mu$m) filter.
The seeing was typically in the range of 0.3--0.5\,arcsec (FWHM; full width-of half maximum) in the mid-infrared, according to observations of calibration stars.
The airmass was between 1.41 and 1.56.
The pixel scale is 0.04\,arcsec, using regular imaging mode, with a field of view of 38$\times$38\,arcsec$^2$.
The standard chop-nod technique was used to subtract sky and instrumental backgrounds, with chopping frequency of 3.99\,Hz, while the nodding  was east-to-west parallel to the chopping direction. 
The amplitude of both the chop and nod throws was 8\,arcsec. 
Although VISIR can offer jittering mode, we used non jittering mode, in order to avoid difficulty in stacking jittered frames.
The total integration time 
was 6450\,sec on source.
Six stars 
(HD41047,
HD49968,
HD35536,
HD39523,
HD47667 and
HD32820)
were used for flux calibration, and their fluxes were estimated from \citet{1999AJ....117.1864C}'s flux calibration list.
The data on February 15th and 17th were taken during clear conditions with high humidity ($>$ 50\%),  thus flux density measurements from those nights seem to suffer from larger uncertainties.

VISIR also obtained Q1 (17.65\,$\mu$m) images, with program IDs of 298.D-5023(A) and 0102.D-0245(A) in 2017 and 2018.
The optical seeing was 0.4--1.5\,arcsec, typically, occasionally reaching 2.0\,arcsec. 
Unless the mid-infrared seeing is really poor, $Q$-band angular resolution is determined by the 8.2-metre telescope's diffraction limit of 0.54" \citep[the diameter of the first ring with `null' signal; ][]{Reunanen.2010}, unlike the normally seeing-limited resolution at $N$-band.
 Indeed, the FWHM of the photometric calibration source was recorded as 0.4--0.5\,arcsec.
 The airmass was 1.40--1.48.
We measure the total flux of these Q1 images for 2017 and 2018, separately, but for our figures in this paper, we combined the data from these two years, in order to increase the image quality.
The pixel scale, field of view and chop-nod settings were the same for B10.7 observations.
There were in total 8 observing blocks, and data from 5 of these were used, as the remaining 3 had poor quality.
That resulted in the total integration time of 11250\,sec on source.
Five flux calibration stars 
(HD39523, HD22663, HD41047 
HD23319 and HD28413) 
were used.


Four years later, VISIR revisited SN\,1987A, imaging in the B10.7 band in January and March 2022 under the program  106.D-2177.
The images were taken at three separate days, 
2022 January 31st (day 12,760),
2022 March 1st (day 12,789) and 
2022 March 26th (day 12,814).
The optical seeing varied between 0.4--2.2\,arcsec on January 31st, whereas it was slightly better on other two days
(0.45--1.73\,arcsec on  March 1st, and 0.4--1.8\,arcsec on March 26th).
Five different calibration sources were used on January 31st (HD026967, HD035536, HD082660, HD041047  and HD083425 ),
while on March 1st, HD041047  and HD055865 were used, and HD112213  and HD114326  on March 26th.
The total integration time was 10,900\,sec on source.

All the VISIR data were reduced, using ESO reduction package, {\sc ESOReflex} \citep{2013A&A...559A..96F}.
The final images were smoothed with a 2-pixel Gaussian filter, in order to increase the signal to noise ratio.

\begin{figure}
	 \includegraphics[width=8cm, trim={2.7cm 1.2cm 1.95cm 0},clip]{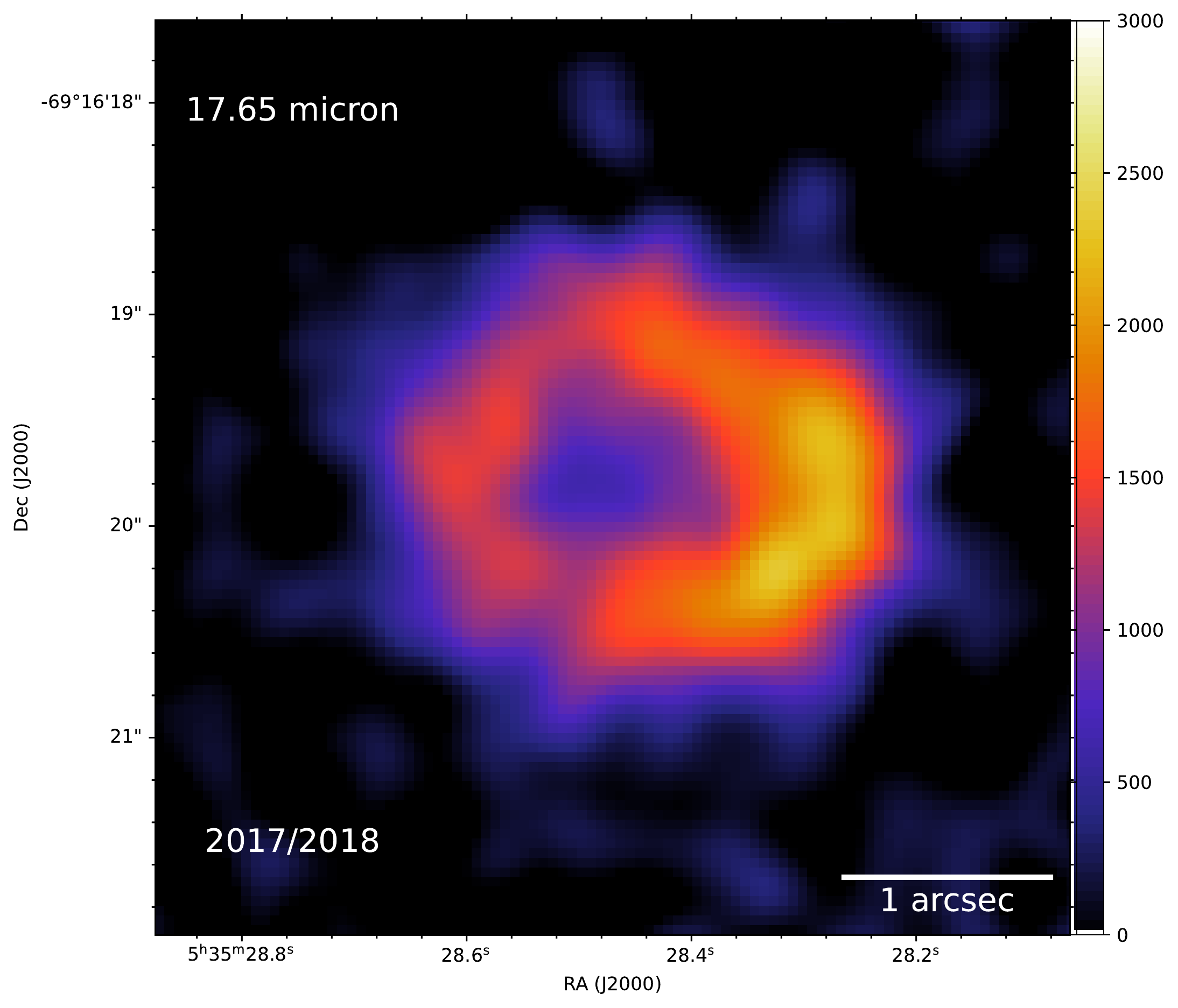}
     \caption{VISIR image of SN\,1987A in the Q1 band (17.65\,$\mu$m), combined from  2017 and 2018 observations.}
        \label{fig:visir_images}
\end{figure}

Figure\,\ref{fig:images} includes  VISIR images of SN\,1987A taken in the B10.7 in 2017  and  in 2022.
Figure\,\ref{fig:visir_images} shows a Q1 band VISIR image of SN\,1987A taken  in 2017 and 2018.
These images clearly show MIR emission from the ring, with the brightest areas being on the south-west side of the ring.
In the B10.7 band, individual clumps within the ring are marginally  resolved.

\citet{Bouchet:2004bs} reported an unresolved central source in their 10\,$\mu$m image in 2003 (also seen in Figure\,\ref{fig:images}). 
That source was not detected in the subsequent 11\,$\mu$m image in 2005 \citep{Bouchet:2006p2168}.
There is a potential point source at the centre of the ring in 10.65\,$\mu$m image in 2017 and 2022 (Fig.\,\ref{fig:images}).
Whether this point source is a real detection or not is unclear, as it is close to the noise level, generated by imperfect reconstruction of the point spread function.
If this point source detection is real, it could be due to ejecta dust \citep{Bouchet:2006p2168}, but also related with the compact source \citep{Alp.2018, Page.202053j}. 
The definitive existence (or not) of this central point source awaits JWST.


\begin{center}
\begin{table*}
\caption{Observing log \label{observing_log} }
\begin{filecontents*}{observing_log.csv}
Instrument,Program ID,Date,Day,Filter,Filter Center,Filter Width,Flux (mJy),Flux error
Gemini-South/T-ReCS,,2003-10-04,6067, $N$ ,10.36,2.89,9.9,1.5
Gemini-South/T-ReCS,,2005-01-07,6526,Si-5,11.66,1.13,18.4,1.2
Gemini-South/T-ReCS,,2005-02-01,6552,Qa ,18.30,1.51,53.4,9.0
Gemini-South/T-ReCS,GS-2004A-Q-3,2005-11-04,7194,Si-5,11.66,1.13,20,2
Gemini-South/T-ReCS,GS-2006B-Q-3,2006-12-22,7242,$N$,10.36,2.89,19,2
VLT/VISIR,080.D-0105,2007-10-07,7533,PAH2,11.88,0.37,48,7
VLT/VISIR,086.D-0192,2010-12-27,8708,B10.7,10.65,1.37,35,6
Gemini-South/T-ReCS,GS-2012B-Q-90,2012-10-25,9376,N,10.36,2.89,38,4
{\em SOFIA/FORCAST},04$_-$0016,2016-07-11,10732,FOR$_{-}$F111,11.2,2.7,45,7
,,,,FOR$_{-}$F197,19.7,5.5,88,9
,,,,FOR$_{-}$F315,31.5,5.7,105,14
VLT/VISIR,298.D-5023,2017-02-15  \& 02-17 \& 03-13,10952--10976,B10.7,10.65,1.37,(26),(5)
,,2017-09-13,11160--11181,Q1,17.65,0.83,114,40
 {\em SOFIA/FORCAST},05$_-$0050,2017-08-03,11119,FOR$_{-}$F197,19.7,5.5,82,11
,,,,FOR$_{-}$F315,31.5,5.7,113,19
%VLT/VISIR,298.D-5023,2017-02-17,,B8.7,~8.92,0.97,,
VLT/VISIR,102.D-0245,2018-12-03,11606--11611,Q1,17.65,0.83,128,16
 {\em SOFIA/FORCAST},07$_-$0064,2019-07-11,11826,FOR$_{-}$F197,19.7,5.5,104,9
,,,,FOR$_{-}$F315,31.5,5.7,76,15
VLT/VISIR,106.D-2177,2022-01-31 \& 03-01 \& 03-26,12760--12814,B10.7,10.65,1.37,42,13
\end{filecontents*}
\csvreader[tabular= l l l c c ll r{@}{$\pm$}l,
				table head=\hline Telescope/Instrument & Program ID & Date & Day & Filter & $\lambda_{\rm eff}$ & $\Delta \lambda$ & \multicolumn{2}{c}{Flux (mJy)}  \\ \hline\hline, 
				table foot=\hline ] 
				{observing_log.csv}
				{} 
				{\csvcoli & \csvcolii  & \csvcoliii  & \csvcoliv & \csvcolv & \csvcolvi & \csvcolvii & \csvcolviii  & \csvcolix } 
\\
\end{table*}
\end{center}


\subsection{SOFIA observations in 2017 and 2019}

Following the Cycle 4 observations in 2016 \citep{Matsuura.20185p8},
the NASA {\it Stratospheric Observatory For Infrared Astronomy} \citep[{\it SOFIA}; ][]{2012SPIE.8444E..10Y} observed SN\,1987A during Cycle 5 in 2017. 
{\it SOFIA}'s FORCAST instrument \citep{Herter:2012hv} obtained photometry during a southern hemisphere campaign with the aircraft temporarily based in Christchurch, New Zealand. 
FORCAST is equipped with two cameras, a short wavelength camera, operating at 5--25\,$\mu$m, and a long wavelength camera, operating at 25--40\,$\mu$m, both of which have a pixel scale of 0.768''\,pixel$^{-1}$. 
We used the dual channel mode of FORCAST, that employs a dichroic to allow imaging in both cameras simultaneously. 
Observations of SN\,1987A were taken on two separate flights (Flight 426 and 427).
On 2017-08-06, FOR$_{-}$F112 (centred at 11.2\,$\mu$m) and  FOR$_{-}$F315 (centred at 31.5\,$\mu$m) filters were used on Flight 427 on 2017-08-06 (11,122 days since the explosion), 
while FOR$_{-}$F197 (centred at 19.7\,$\mu$m) and FOR$_{-}$F315 were used for the Flight 426 on the 2017-08-03 (day 11,119).
The total exposure times on source were 3181\,sec for FOR$_{-}$F112, 2274\,sec for FOR$_{-}$F197, 3486+2686\,sec for FOR$_{-}$F315, respectively.
Observations were performed using the ``Nod-Match-Chop'' mode, i.e. the same method employed by the VLT for the VISIR observations mentioned earlier,
and were configured to have 45'' East-West chop and nod throws. 
Unfortunately, SN 1987A was not detected in the FOR$_{-}$F112 band.

  {\it SOFIA} further observed SN\,1987A with FORCAST  in 2019 on July 1st, and July 8th--11th. 
Similar to previous flights, ``Nod-Match-Chop'' mode was used. During these flights (Flight 588 and 592--595), FOR$_{-}$F197 and FOR$_{-}$F315 were observed separately, without the dichroic.
The total exposure time was 15016\,s for FOR$_{-}$F197 and 16070\,s for FOR$_{-}$F315.

\subsection{Archival data}

SN 1987A was monitored in the mid-infrared by Gemini-South/T-ReCS \citep{2010SPIE.7737E..2FL} and the observations up to 2003 have been reported by \citet{Bouchet:2006p2168}.
Fluxes in Table\,\ref{observing_log} are taken from \citet{Bouchet:2006p2168}.
Further images were taken by Gemini-South T-ReCS and  VLT/VISIR from 2005 to 2012, and their fluxes are also listed in Table\,\ref{observing_log}.
VISIR underwent detector upgrade in mid 2010's, and the VISIR images taken in 2007 and 2010 before this upgrade had a pixel scale of 0.075".


{\em SOFIA} observed SN\,1987A with {\em FORCAST} in 2016, and the fluxes are taken from \citet{Matsuura.20185p8}.

\section{Results}

\subsection{Time evolution in the mid-infrared images with comparison to optical and X-ray images}

The {\it Hubble Space Telescope} ({\it HST}) $R$-band images in  Figure\,\ref{fig:images}, taken from \citet{Larsson2019},
 demonstrate the structures and evolution of the remnant of SN\,1987A.
The bright ring, which is often called the `inner ring' or `equatorial ring', is one of the circumstellar rings found in SN\,1987A, with the other ring being  fainter and farther (up to  $\sim$5" in diameter) `outer rings' \citep{1996ApJ...459L..17P}.
The inner ring is composed of about twenty clumps within, while the ejecta appear as a central keyhole-shaped cloud, expanding in time. 

\subsubsection{Evolution of brightness distribution within the ring} \label{sect-brightness}

The {\it HST} monitoring program observed that as the ejecta expanded, the ring brightened until $\sim$2011, 
and then began getting fainter \citep[Figure\,\ref{fig:images}; ][]{Fransson:2015gp}.
At $R$-band, in 2003 the ring was relatively faint, and gradually increased its brightness in time.
The ring was brightest at the north-east and south-west  \citep{Fransson:2015gp}.
In 2005 (day$\sim$7000),  some of the knots in the north-east side of the ring started fading  \citep{Fransson:2015gp, Larsson2019}.
The peak brightness shifted from the east  to the west side of the ring in 2006 \citep{Fransson:2015gp, Larsson2019}.
The brightness on the west side continued to increase till 2011, and then started fading, too \citep{Larsson2019}.

  The second row of Figure\,\ref{fig:images} shows the mid-IR (mid-IR) time evolution of SN\,1987A images.
The time evolution at MIR wavelengths follows more or less similar trends to the ring evolution in {\it HST} images.
Initially only the east side of the ring was visible in the 2003 image. 
As the blast waves progressed, the north-east side of the ring brightened, first in the mid 2000's.
Eventually, the brightness of the west side of the ring caught up with the east side in 2007, and in 2017 the west side of the ring is the brightest.

In the X-ray, {\it Chandra} images at 0.3--8\,keV show that the east side of the ring was bright until 2007 (day\,$\sim$7000) \citep{Frank:2016ka}.
Since 2009,  the west side of the ring is dominating the brightness.
Unlike {\it HST} images, the total brightness at X-ray wavelengths continued to increase until 2009 (day$\sim$9000).
In the 0.5--8\,keV and 0.5--2\,keV bands, the total brightness plateaued, while the 3--8\,keV brightness continued to increase beyond 2009  \citep{Frank:2016ka}.

\begin{figure}
	\includegraphics[width=8.5cm]{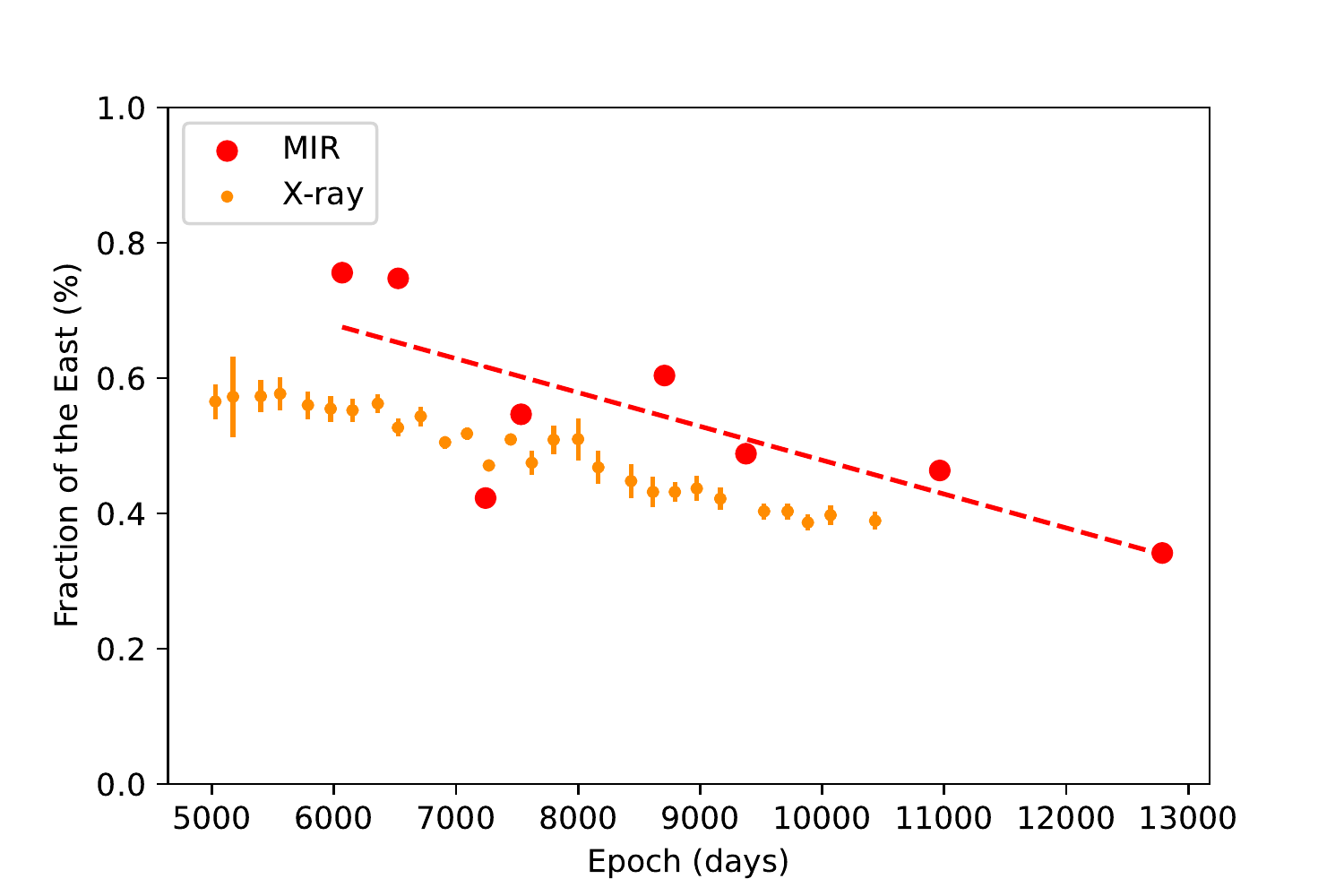}
    \caption{The fraction of the MIR emission originating from the east half of the ring with respect to the total emission at 10--12\,$\mu$m.
    The dashed line is a linear fit to the measured fractions.
    The time variation of the fraction in X-ray  \citep{Frank:2016ka} is also plotted.
    Both X-ray and MIR show declining trend in the fraction emitted from the east half, however, the fraction tends to be higher in  MIR than  X-ray.
            \label{fig:east-west}}
\end{figure}

Fig.\,\ref{fig:east-west} shows the time evolution of east-west ratio of the emission within the ring at 10--12\,$\mu$m.
The diagram follows the fraction of the emission coming from the east side of the ring, with respect to the total emission.
The uncertainties, estimated by shifting the east-west border by one pixel, were typically less than a few per cent.
The ratio taken in 2006 (day 7242)  is a low outlier in this figure, and it appears that this data set seems to have suffered from negative backgrounds in the surroundings.
The east side of the ring produces nearly 80\,\% of the total brightness in day$\sim$6000, and this fraction declines in time, down to $\sim$40\,\% in day 12,800.

Although a general decreasing trend of the east fraction is consistent with that found in X-ray images at 0.3--8\,keV \citep{Frank:2016ka},
the actual IR values tend to be higher.
Around day$\sim$6,000, 
the IR fraction from the east was nearly 80 per cent, whereas for X-rays it was only 57 per cent.
The fraction dropped to only about 50\,\% at day 10,000, while the X-ray fraction from the east was 38\,\%.
The IR emission might have a delayed response time, after the shock hit the ring, traced by X-ray emission.

\subsubsection{Size of the ring}

\begin{figure}
	\includegraphics[width=8.5cm]{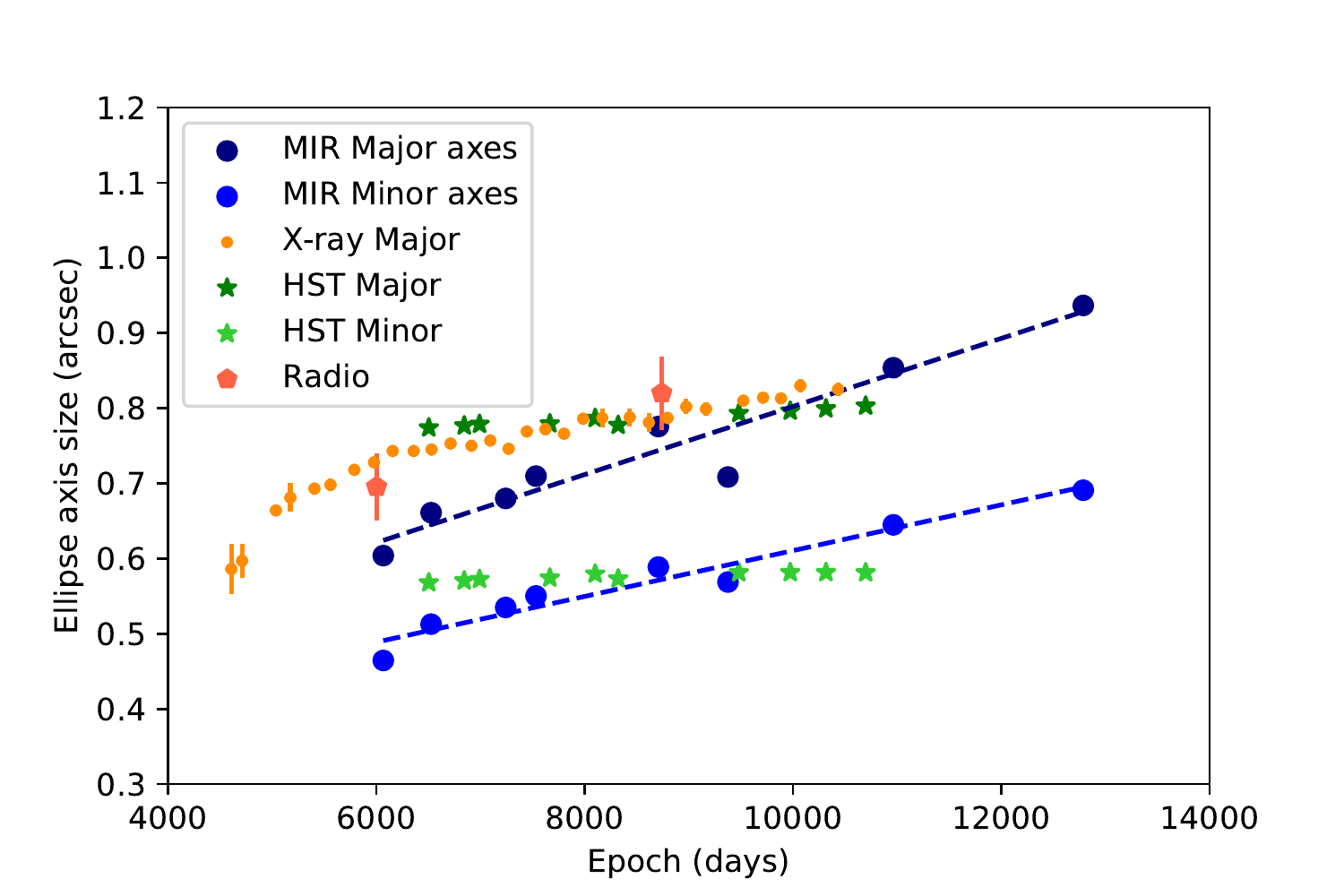}
    \caption{The  major and minor axes of ellipse fitting to the 10--12\,$\mu$m MIR ring images,  as a function of time.
    The ellipse is expanding.
    As a comparison, X-ray \citep{Frank:2016ka} {\it HST} optical ellipse sizes and the 18\,GHz 
    synchrotron east-west radii, which are equivalent to the ellipse major axes, from \citet{Zanardo:2013fn},
    are also plotted. The X-ray size increased initially but flattened recently. The optical size remains almost constant.
    The radio size seems to follow the X-ray trend.
            \label{fig:ellipse}}
\end{figure}

Figure\,\ref{fig:images}  suggests that the mid-infrared diameter of the ring may be slightly larger in 2017 (day 10952--10976) and in 2022 (day 12760--12814), compared with 2005 (day 7194).
Figure\,\ref{fig:ellipse} shows that the major and minor axes measured in mid-infrared images are increasing in time.
The fitting used 2D Gaussian fitting, utilising {\sc mpfit2dfun} in {\sc IDL} \citep{2009ASPC..411..251M}.
The B10.7-band angular resolutions are limited by the seeing, and the FWHMs of the point-spread functions (PSFs) of the calibration stars were recorded at 0.3--0.5\,arcsec in 2017 and 0.3--0.4\,arcsec in 2022.
Given this angular resolution, the increasing trend over 7,000 days is likely to be real.
Although we have to be careful about unexpected image distortions, caused by  chop-nod observations,
the ring seems to expand over time in the MIR images.
The lines in Figure\,\ref{fig:ellipse} are fits to the measured ellipse sizes, corresponding to expansion at 3920\,km\,s$^{-1}$ for the major axis, and 2636\,km\,s$^{-1}$ at 50\,kpc.
The inclination angle of the ring was estimated to be 38--45$^{\circ}$ \citep{1995ApJ...452..680B, Larsson2019}, and our measured ratio of major and minor axes is consistent with this inclination angle..

In the optical, there is nearly no expansion found.
{\it HST}  images recorded 
the diameter of the ring could be expanding by merely 0.02--0.04\,arcsec  between day 7000 and 9000 \citep{Fransson:2015gp}.
 We measured {\it HST} ellipse sizes before day 7000 and beyond 9000 (Fig.\,\ref{fig:ellipse}), and even in the longer time span, the size remains almost constant.

On the other hand, the X-ray radius has been reported to have increased from 0.728\,arcsec to 0.830\,arcsec from day 5978 (in 2003) to day 10073 (in 2014) \citep{Frank:2016ka}, and their measurments are replicated in Figure\,\ref{fig:ellipse}. 
These reported X-ray sizes are more or less consistent with the MIR estimates of the major axes, but slightly smaller.
The X-ray has provided more frequent measurements, and up to day $\sim$6500, the expansion of the radius at 0.5--2\,keV was fitted by 6711$\pm$787\,km\,s$^{-1}$ and later epochs by 1854$\pm$101\,km\,s$^{-1}$ \citep{Frank:2016ka}. 
The MIR expansion rate was 3920\,km\,s$^{-1}$ for the major axis, which was  faster than the X-ray 0.5--2\,keV  band until recently. 
The higher energy X-ray band at 2--10\,keV continues to expand at a higher rate of 3071$\pm$299\,km\,s$^{-1}$ \citep{Frank:2016ka}, closer to MIR velocities.
The radial expansion velocity was also measured in synchrotron radiation at radio frequencies, and
the  expansion velocity measured at 9\,GHz \citep{Ng:2013bt, Zanardo:2014gu} continued to be 3890$\pm$50\,km\,s$^{-1}$ \citep{Frank:2016ka}, which is close to the MIR measurement.

\subsection{Dust temperature indicator}

\begin{figure}
	\includegraphics[width=8cm]{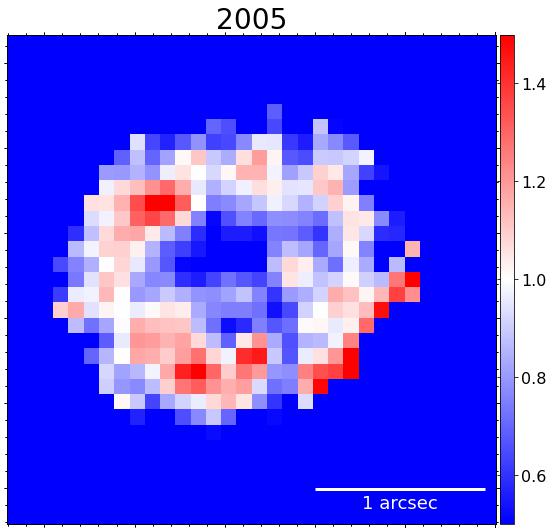}
	\includegraphics[width=8cm]{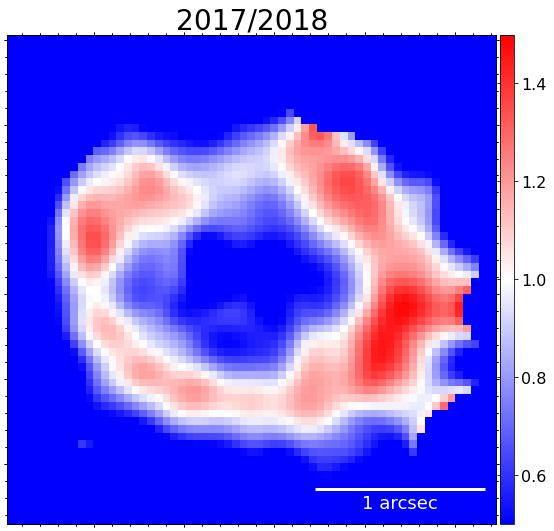}
    \caption{ The relative brightness ratios of $I_{\nu}(\lambda=11.88)/I_{\nu}(\lambda=18.30)$  in 2005 (top) and $I_{\nu}(\lambda=10.7)/I_{\nu}(\lambda=17.65)$   in 2017 and 2018 (bottom). 
    The ratios are scaled so that the mean values of surface brightnesses become unity.
    Higher values of ratios, indicating higher dust temperatures within the ring, are found in the north-east in 2005, while they are located in the south-west in 2017/2018
    \label{fig:dust_tmp}}
\end{figure}

 The ratio of two MIR band images gives a measure of the dust temperature distributions within the ring.
Fig.\,\ref{fig:dust_tmp} shows the relative brightness ($I_{\nu}$) ratios of $I_{\nu}(\lambda=11.88)/I_{\nu}(\lambda=18.30)$ in 2005 (top) and $I_{\nu}(\lambda=10.7)/I_{\nu}(\lambda=17.65)$  in 2017 and 2018 (bottom). 
Unfortunately, there was some issue in the absolute flux calibration of $I_{\nu}(\lambda=10.7)$ in 2017, so that we have taken the median of the surface brightnesses across the ring, and these ratios indicate excess above/below the median values.
Although these ratios are unable to give the actual dust temperatures, they  are indicative of warmer and colder dust temperatures within the ring at a given time.

In 2005, the north-east part of the ring shows relatively higher ratios, indicating higher dust temperatures.
That is consistent with the location of the brightest part of the ring in the MIR image (Fig.\,\ref{fig:images}).
In 2017, the brightest part of the the ring is shifted towards the south-west (Fig.\,\ref{fig:images}), and that part of the ring also has the highest ratio of $I_{\nu}(\lambda=10.7)/I_{\nu}(\lambda=17.65)$, indicating the highest dust temperature at this location.

Brightness ratio maps show that the shift of brightness peak from north-east to south-west is mainly caused by the change of the dust temperatures.
The brightness of MIR dust emission depends on both dust temperature and column density along the line of the site.
When it is optically thin ($\tau(\lambda)\ll1$) 
\footnote{The optical depth, calculated from $M_d$ from Sect.\ref{sec-SED}, 
with an assumption of the clump thickness of 0.05" and an ellipse size from 
Fig.\,\ref{fig:ellipse} is $\tau_d$=0.002 at 10\,$\mu$m, i.e., optically thin on day 10,964.}, 
the brightness of the dust emission reaches
$I_\lambda= \tau_d(\lambda) B_\lambda (\lambda, T_d)$ \citep{1978ppim.book.....S, 1986rpa..book.....R},
where $\tau_d(\lambda)$ is the dust optical depth at wavelength $\lambda$, and $B_\lambda (\lambda, T_d)$ is the Planck function  at the dust temperature $T_d$ \citep{Gordon:2014bya}.
The optical depth $\tau_d(\lambda)$ is represented by  $\kappa(\lambda)\Sigma_d$, where $\kappa(\lambda)$ is the dust absorption cross section and $\Sigma_d$ is the dust surface mass density  \citep{1978ppim.book.....S, Gordon:2014bya}.
The ratio of brightnesses at a given time does cancel out $\Sigma_d$, hence, this value only depends on $T_d$.
The time variation of relative brightness shift from the east to west (Fig.\,\ref{fig:images}) is influenced by the dust temperature $T_d$.
The $T_d$ was higher in the east in 2005, while it is higher in the west in 2017/2018.



\subsection{Time evolution of the MIR spectral energy distributions} \label{sec-SED}

\begin{figure}
	\includegraphics[width=8.5cm]{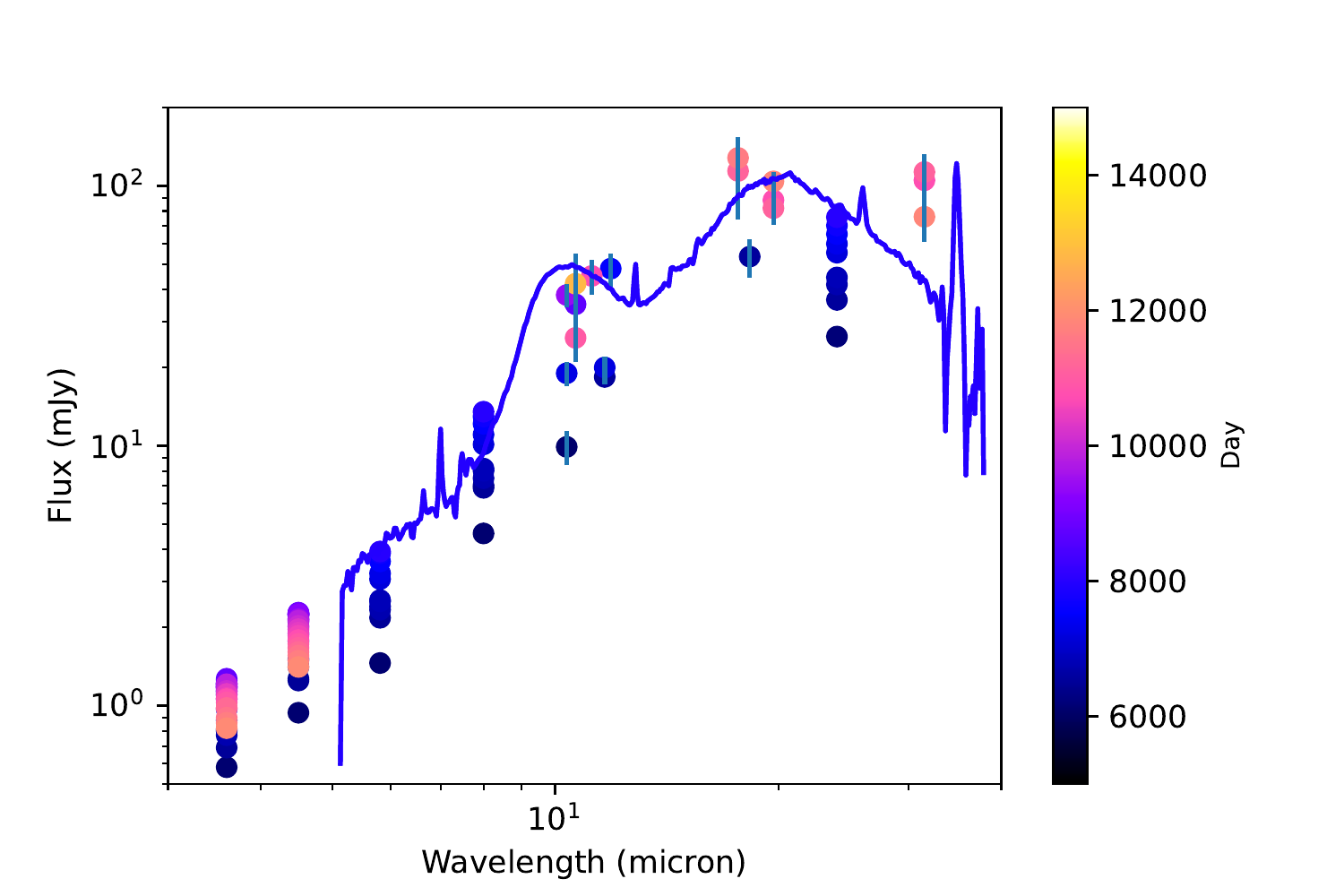}
    \caption{ The time evolution of the MIR spectral energy distributions. The colours of  the plotted points indicates the dates of observations, shown in the colour scale bar. The plot includes {\it Spitzer} IRS 5--28\,$\mu$m spectrum, with corresponding colour for the date observations.
    \label{fig:sed-ini}}
\end{figure}

Fig.\,\ref{fig:sed-ini} shows the time variation of the spectral energy distributions (SEDs).
The colour of the plotting points indicate the observed day since the explosion.
The majority of the photometry data points are from {\it Spitzer Space Telescope} monitoring programs
\citep{Dwek:2010kv, Arendt:2016ds}.
These include 3.6, 4.5, 5.8, 8.0 and 24\,$\mu$m photometric data points, as well as 5--35\,$\mu$m Spitzer/IRS spectra.
The Spitzer/IRS spectrum shows that photometry at $\sim$10 and $\sim$20\,$\mu$m are strongly affected by the broad silicate features.
Although {\it Spitzer} took  spectra at multiple epochs, only the last spectrum at day 7954 is plotted in Fig.\,\ref{fig:sed-ini}.
 {\it Spitzer} monitoring at 3.6 and 4.5\,$\mu$m continued after {\it Spitzer}'s  helium coolant was exhausted, showing that their fluxes reached a peak at about day 8500--9000 in 2009--2010 \citep{Arendt2020}.
Other {\it Spitzer} bands were last observed on day 7975--7983, just before these bands would have peaked in brightness \citep{Arendt2020}.

All ground-based photometry at 10--12\,$\mu$m after day 10,000 falls below the day 7954 {\it Spitzer} spectra.
At 18--20\,$\mu$m and 31.5\,$\mu$m such a trend is unclear.
Within the uncertainties, 18--20\,$\mu$m fluxes after day 10,000 are more or less comparable to the flux measured by {\it Spitzer} on day 7954.
The {\it SOFIA} 31.5\,$\mu$m fluxes are higher than day 7954 {\it Spitzer} spectra at the same wavelength, as reported  by \citet{Matsuura.20185p8}.

The brightness ratios of the N- and Q-band images demonstrate the spatial variations of the temperature within the ring
(Fig.\,\ref{fig:dust_tmp}). 
We evaluate whether these variations may be reflected in the overall SED, by fitting the SED with a modified blackbody, representing the dust emission.

The flux density $F_{\nu}$  with frequency grid $\nu$  from a dust mass ($M_d$) can be characterised by a modified blackbody of the form of
\begin{equation}
  F_{\nu}(\lambda) = M_{d} \frac{4 \kappa(\lambda) \pi B_{\nu}(\lambda, T_d)}{4 \pi D^2}, 
   \label{eq:dust}
\end{equation}
where $M_d$ is the dust mass,
$B_\nu(\lambda, T_d)$ is the Planck function,
and $T_d$ is the dust temperature  \citep{Hildebrand:1983tm}.
This modified blackbody assumes that the dust emission is optically thin.
$D$  is the distance to the LMC, adopted to be 50\,kpc. 
$\kappa(\lambda, a)$ is the dust mass absorption coefficient,
$  \kappa(\lambda, a) = 3 Q(\lambda)/4 \rho  a$,
where $\rho$ is the mass density of the dust grains, $Q(\lambda)$ is the dust emissivity at the wavelength
$\lambda$, and $a$ is the grain size.
In this SED fit, we assumed a single grain size of 0.1\,$\mu$m.
 This grain size is relatively large within the typical grain size distributions adopted in the ISM  \citep[50\,\AA--0.25\,$\mu$m; ][]{Weingartner:2001p3411}, however, slightly smaller but similar to the past estimate of dust in the ring  \citep[$>$0.2\,$\mu$m; ][]{Dwek:2010kv}.
This assumption affects estimated $T_d$ and $M_d$, but not the total IR luminosity.

%
First, we fitted the data near the peak of the silicate feature with a single dust component (case 1).
Silicate dust emissivity is calculated from \citet{Draine:1984p25590}.
For day 7954, the Spitzer IRS spectrum from 5\,$\mu$m to 33\,$\mu$m was taken into account.
For day$>$10,000, we used all photometry, including {\it SOFIA} 31.5\,$\mu$m for case 1.

Case 2 uses two component fits for hot and warm dust, following \citet{Dwek:2010kv}. 
The warm dust component represents silicate emission around 8--20\,$\mu$m, similar to case 1.
There is an additional emission component to fit {\it Spitzer} 3.6 and 4.5\,$\mu$m data points, and that contributes to the underlying emission at longer wavelengths \citep{Dwek:2010kv}.
Although the exact nature of this emission component is unknown, we call it `hot' dust, and fit with amorphous carbon \citep{Zubko:1996p29442}. 
\citet{Dwek:2010kv} fitted this component with alternative dust compositions, such as Fe, Fe$_2$O$_4$, FeS, but in this work, we tested only with amorphous carbon, as the exact composition of dust is unknown only with two bands.
Although we tried to fit the hot component with silicates, the resultant fit had a far higher $\chi^2$ value, so that we excluded it from further analyses.

There is additionally a cold ejecta dust component, representing far-infrared emission at 70\,$\mu$m and longer wavelengths
\citep{Indebetouw:2014bt, Matsuura:2015kn, Matsuura.20185p8, Cigan:2019cl}.
At the moment, this component is not included in the analysis, because the mid-infrared dust emission is dominated by the ring dust.

Best fitted parameters were searched, using the {\sc IDL} version of the {\sc amoeba} function \citep{1989nrca.book.....P},
and the uncertainties of parameters were evaluated by Monte Carlo method, using the {\sc IDL} version of {\sc randomn} function \citep{1989nrca.book.....P}.
Table\,\ref{model_fit} summarises the parameters derived from the fitted results.

The dust temperatures ($T_d$) of a single component fit (case 1) suggest that the overall temperature might have dropped from 185.9$\pm$0.1\,K to 154$\pm$5\,K from day$\sim$7954 to day $>$10,000.
The very small uncertainty of $\pm$0.1\,K on early days is due to the very small uncertainties of {\it Spitzer} IRS spectra, which are propagated into the temperature uncertainties. The uncertainties do not include any systematic errors related to the either the flux measurements (e.g. calibration) or the choice of $\kappa(\lambda)$ for the model.
However, the warm dust temperatures  for a two-component fit (case 2) remain more or less the same.
Indeed, the case 2 hot component fit has smaller dust temperature uncertainties at day$>$10,000 than the day$\sim$7954 fit, but large dust mass uncertainties, with no constraints, suggesting a shortcoming of our simple fit.
Therefore, even though spatially resolved images indicate the change of dust temperature at different parts within the ring, the SED is rather insensitive to these changes, because the SED contains both increasing and decreasing small-scale temperature changes. 

\begin{figure*}
  \begin{minipage}[c]{1\textwidth}
	\includegraphics[width=9cm]{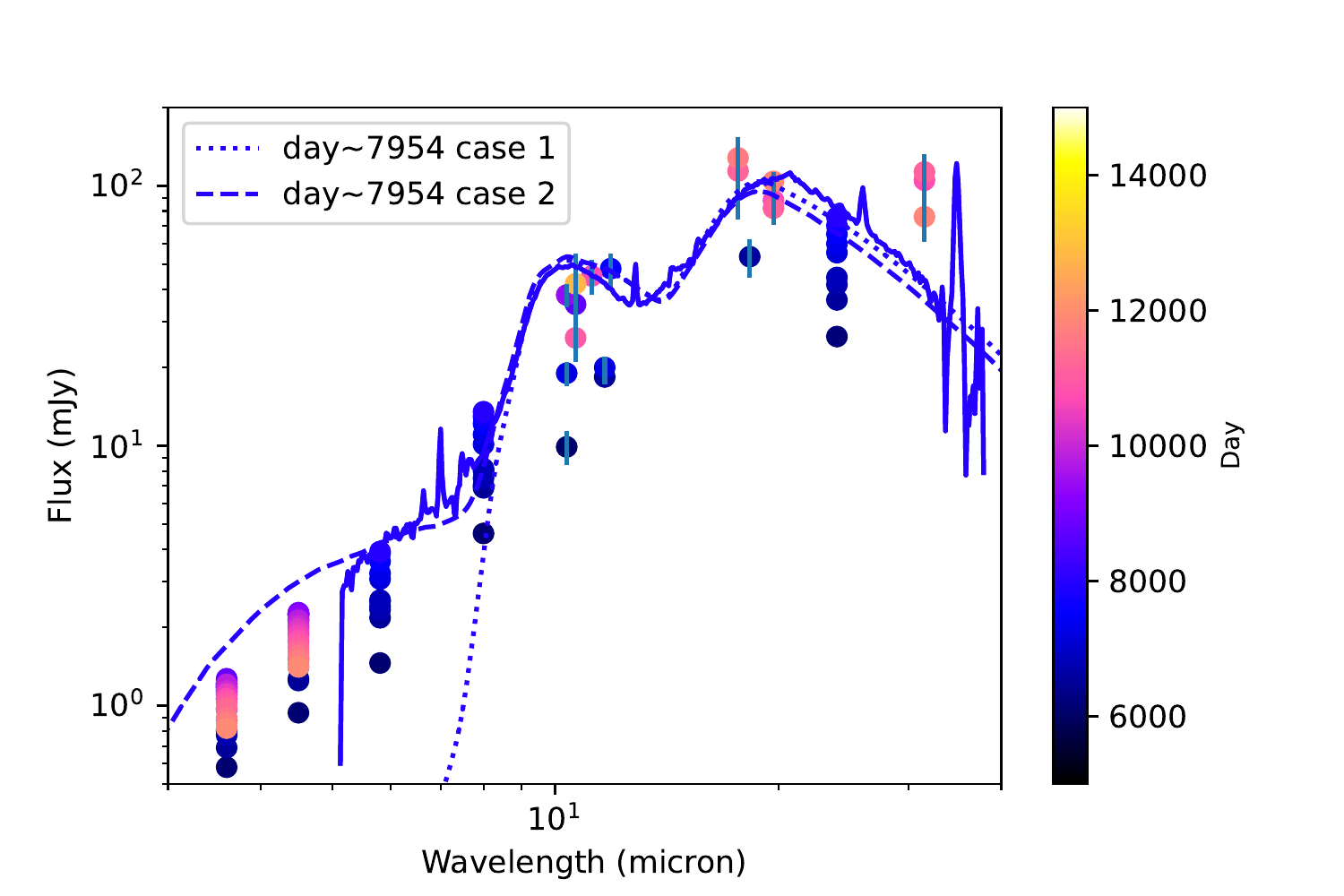}
	\includegraphics[width=9cm]{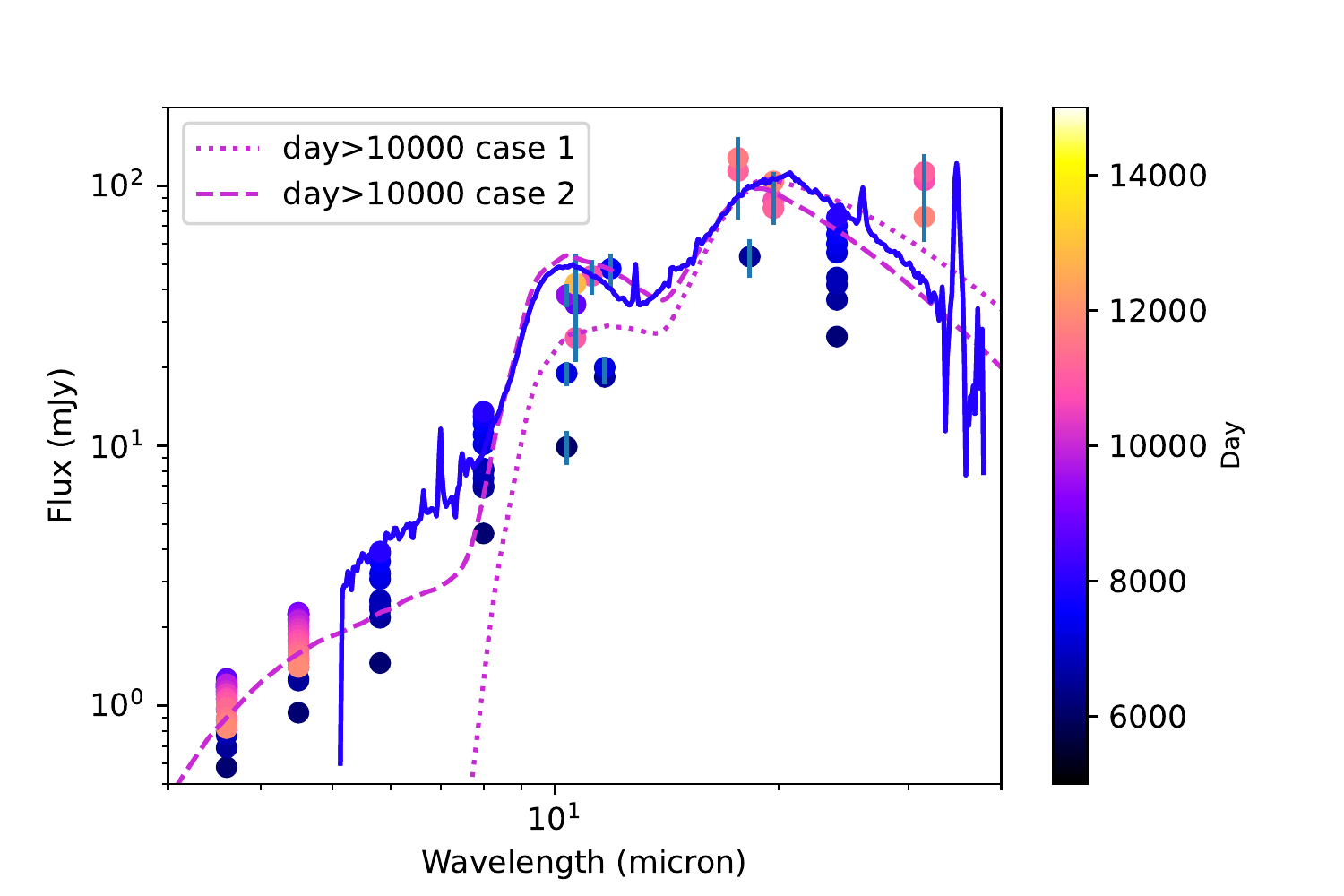}
      \end{minipage}\hfill
    \caption{ Dust model fittings to the photometry and Spitzer IRS spectra at day $\sim$7,954 (left) and photometry at day$\sim$10,000 (right).
    Note that all data points in Fig.\,\ref{fig:sed-ini} are plotted in both panels, so that the time variations can be found. The fitted results have little difference in the flux levels at 10 and 19\,$\mu$m.
    \label{fig:sed}}
\end{figure*}

\begin{table*}
	\centering
	\caption{The results of modified blackbody fits to the IR SEDs at day $\sim$7954 and day$>$10000. Case 1 is for a single component dust emission and case 2 is for two component dust emissions.}
	\label{model_fit}
	\begin{tabular}{lll c   c} 
		\hline \hline
		& & & day$\sim$7954  							&  day$>$10000 \\
		\hline 
Case 1	& Data && IRS only								& $>$8\,$\mu$m only\\
		&& Dust & Silicate								& Silicate \\
		&& $M_d$ (\Msun) &  (1.110$\pm$0.003)$\times10^{-5}$ 	&  (2.6$\pm$1.3)$\times10^{-5}$ 	\\
		&& $T_d$  (K)        &  184.9$\pm$0.1				& 154$\pm$5\\ \hline
Case 2     & Data && 3.5, 4.6, 5.8, 8.0\,$\mu$m \& IRS			& 3.6, 4.5\,$\mu$m \& $>$8\,$\mu$m \\
		& {\bf Hot} & Dust & Amorphous C 					& Amorphous C \\
		&& $M_d$ (\Msun) &  (1.31$\pm$0.07)$\times10^{-8}$ 	&  (0.7$\pm$0.7)$\times10^{-8}$ 	\\
		&& $T_d$  (K)        & 525.6$\pm$0.5 				& 525.40$\pm$0.09\\
		& {\bf Warm} & Dust & Silicate 						& Silicate \\
		&& $M_d$ (\Msun) &   (0.90$\pm$0.03)$\times10^{-5}$	& (0.9$\pm$0.3)$\times10^{-5}$ \\
		&& $T_d$  (K)        & 190.9$\pm$1.0					& 191.3$\pm$0.2\\
		\hline \hline
	\end{tabular}
\end{table*}

\subsection{Light curve analysis}

\begin{figure}
	\includegraphics[width=8cm]{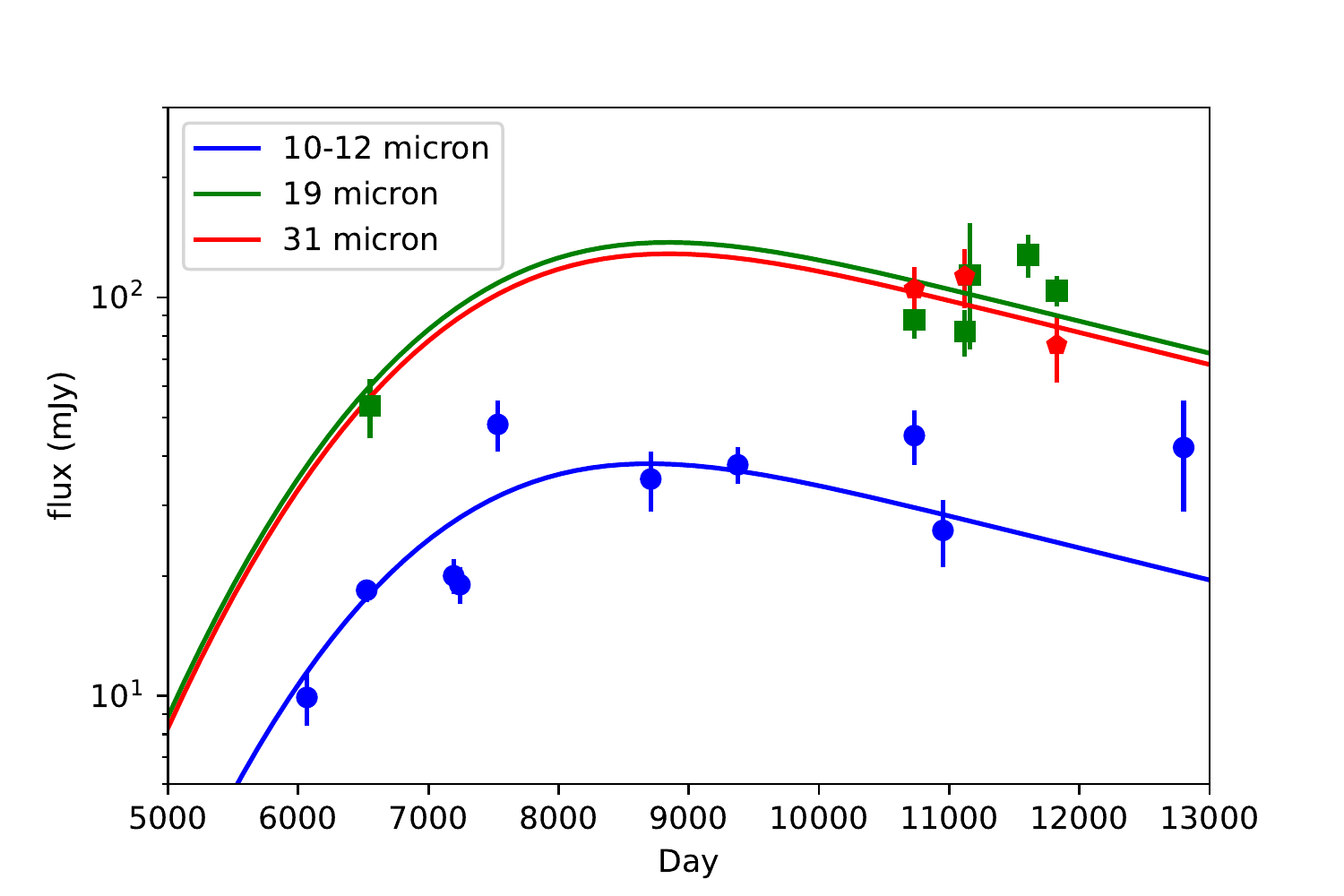}
    \caption{ The time evolution of flux densities at 10--12, 19 and 31\,$\mu$m, assembled from multiple measurements from GEMINI, VLT and {\it SOFIA} (Table\,\ref{observing_log}).
    The light curves, simply scaled from those fitted to the {\it Spitzer}  8 and 24\,$\mu$m flux densities,  can replicate the flux evolutions very well.
    \label{fig:lightcurve}}
\end{figure}

Fig.\,\ref{fig:lightcurve} shows light curves at three different wavelengths.
The measurements at 31\,$\mu$m are solely from {\it SOFIA}, while those at 10--12\,$\mu$m and 19\,$\mu$m were collected from different instruments with slightly different filter systems, so that they are not a completely uniform sample.
Nevertheless, these light curves generally follow those estimated from {\it Spitzer} 3.6 and 4.5\,$\mu$m measurements \citep{Arendt2020}.
\citet{Arendt2020} demonstrated that the 3.6 and 4.5\,$\mu$m light curves are well represented by a convolution of exponential decay and a Gaussian function.
These fittings are extended to the 8 and 24\,$\mu$m {\it Spitzer} bands, too.
The curves in Fig.\,\ref{fig:lightcurve} use the same Gaussian $\sigma$ and exponential $\tau$ parameters as {\it Spitzer}  8 and 24\,$\mu$m but the heights are scaled.
Surprisingly, the 19\,$\mu$m fluxes from GEMINI and ESO/VISIR follow the  {\it Spitzer}   24\,$\mu$m light curve well.
The 10--12\,$\mu$m fluxes follow the scaled {\it Spitzer}   8\,$\mu$m light curve, but have a larger scatter.
This scatter is most likely because these wavelengths cover the silicate features, and depending on the filter transmissions, flux can change substantially.
The light curves demonstrate that there is very little difference in flux level at day$\sim$7,954 and 10,000, as found in Fig.\,\ref{fig:sed}, because the fluxes still increased from day$\sim$7,954, reaching a peak at about day$\sim$8,500, and then decreasing afterwards, coming back to a similar level to that around day$\sim$7,954.

\section{Discussion}

The IR emission from the ring is produced by dust grains that are collisionally heated by the same gas that also gives rise to the X-ray emission. 
This creates an intimate connection between the basic physical properties of the shocked gas and its X-ray and IR emissions. 
A measure of this connection is given by IRX, the IR-to-X-ray flux ratio in the dusty plasma \citep{Dwek:1987p3350}.

Many factors determine the value of IRX: the gas  density and temperature, which determine the collisional heating rate of the dust and the free-free emission from the gas; the dust-to-gas mass ratio, which determines the relative efficiency of dust and atomic processes in the cooling of the shocked gas; and the grain size distribution which, for a given dust-to-gas mass ratio, determines the efficiency in which the grains absorb the energy of the incident electrons \citep{Dwek:1987p3350, Dwek:2010kv, Koo:2016hba}.

At sufficiently high electron temperatures or sufficiently small grain radii, electrons penetrate the grains and the dust temperature becomes independent of plasma temperature, and a good diagnostic of gas density. On the other hand, for sufficiently low gas temperatures or large grain radii the electrons are stopped in the grains and the dust temperature provides a strong constraint on the plasma temperature and density.

 Models of the X-ray emission from the ring \citep{Ravi.2021} suggest that X-ray emission comprises of a soft X-ray emitting component, generated by a $\sim 10^3$~km\,s$^{-1}$ shock propagating through the dense clumps of the ring, and a hard X-ray emitting component, generated by the shock propagating through the lower density inter-clump region and the less dense medium in which the ring is embedded.  
 The soft X-ray component is characterised by a temperature of about 0.8~keV ($T_e \approx 10^7$~K), which is consistent with that behind a transmitted 680~km\,s$^{-1}$ shock through the clumps \citep{Larsson2019}.
 Using optical spectra of the ring,  \citet{Mattila:2010p29482} derived preshock densities between  $\sim$(1--30)$\times 10^3$~cm$^{-3}$ by fitting emission line light curves with photoionisation models.
These plasma  densities and temperature are capable of collisionally heating the silicate dust grains in the ring to the observed temperature of warm $\sim$180\,K
\citep[Figure 6 of ][]{Dwek:2010kv}. However, densities above $2\times 10^4$~cm$^{-3}$ are required to collisionally heat the secondary hot dust component to temperatures of $\sim$400--500\,K.

For the temperatures inferred from the X-ray observations, most electrons are stopped in the grains. 
The electron density required to heat the silicate grains to the observed temperature of 180~K is then linearly proportional to the grain and for a fixed electron temperature is approximately given by  $n_e({\rm  cm}^{-3}) \approx (4\pm 2)\times 10^4\times a$, where $a$ is the grain size in $\mu$m \citep[Figure 6 of][]{Dwek:2010kv}.

 The dust is also destroyed by collisions with the nuclei in the hot plasma. 
 The thermo-kinetic sputtering of the dust, is dominated by collisions with H and He nuclei. In a gas that is shocked to temperatures between (1--10)$\times 10^6$\,K, the lifetime of silicate dust ($t_d$) is about $t_{\rm d} (yrs) = (0.5-1.0)\times10^6 a(\mu m) / n_{\rm ion} ({\rm cm}^{-3})$, where $n_{\rm {ion}}$ is ion density in cm$^{-3}$ 
 \citep{Nozawa:2006p29444, Dwek.19964y}.
Given the constraint imposed by the dust temperature on the relation between dust radius and gas density, and the uncertainties in the relations, we estimate that the dust lifetime in the shocked ring is between 8 to 32 yrs. 

The lifetime of the dust is comparable to the age of the shocked gas. 
Depending on the details of the shock parameters, a steady supply of dust is needed to sustain the observed IR emission over time.  
A steady supply of dust is guaranteed as long as the shock expands into the ring. 
However, eventually, the shocks passes through the ring \citep{Kangas.2021}, halting the supply of new dust grains, and depending on the dust lifetime, and plasma cooling time the IR emission from the ER will decline with time \citep{Arendt:2016ds, Arendt2020}.

The IRX  remained fairly constant at a value of  2.6$\pm$0.5 until day$\sim$8000,  where IR brightness was mainly monitored by  {\it Spitzer} \citep{Dwek:2010kv}.
 After its helium coolant was exhausted, {\it Spitzer} monitoring continued only at 3.6 and 4.5\,$\mu$m, without covering the peak of the IR SED.
Nevertheless,
after day$\sim$8,000, the fluxes at these two bands plateaued, and then started decreasing at about day 9,000
\citep{Arendt:2016ds}.
In contrast, the X-ray brightness at  0.5--2.0\,keV continued to increase \citep{Arendt:2016ds,Frank:2016ka}.
The existing {\it Spitzer} monitoring suggests that the IR/X-ratio seems to decrease after day 9,000.

We evaluate the total IR brightness from our mid-IR SEDs, and test if indeed the IRX has changed after day 9,000.
The IR flux densities between 3.5 and 40\,$\mu$m from the case 2 SED fits (Sect\,\ref{sec-SED})
result in fluxes of (1.62$\pm$0.02)$\times10^{-11}$\,erg\,s$^{-1}$\,cm$^{-2}$ and (1.57$\pm$0.07)$\times10^{-11}$\,erg\,s$^{-1}$\,cm$^{-2}$ at day$\sim$7954 and day$>$10,000, respectively.
There is little change over the time within the uncertainties. 
This is more or less consistent with the light curve analysis in Fig.\,\ref{fig:lightcurve}, when taking into account the uncertainties in the fluxes.

In contrast, the X-ray brightness at 0.5--2.0\,keV continued to increase from 
(5.6$\pm$0.2)$\times10^{-12}$\,erg\,s$^{-1}$\,cm$^{-2}$ to  (7.9$\pm$0.2)$\times10^{-12}$\,erg\,s$^{-1}$\,cm$^{-2}$ between day 8,000 and 10,433 \citep{Frank:2016ka}.
Hence, the IRX decreased from 2.9$\pm$0.1 to 2.0$\pm$0.1 during this time.

The change in IRX coincides with the time when the brightest part of the ring shifted from the east to the west (Sect.\ref{sect-brightness}).
We can assume that the IRX of the ring is dominated by the shocks in the east until day $\sim$7,000, and after day 8,000, the ratio traces more emission from the west side of the ring.


Determining the causes leading to changes in IRX are
complicated, because the degeneracy in many of the physical
parameters determining its value. The most fundamental
parameter determining IRX is the dust-to-gas mass ratio,
which can decease with time because of the effect of grain
destruction. 
The decrease in grains radius due to sputtering can be written as 
$\Delta a (\mu{\rm m}) = n_{\rm ion} ({\rm cm}^{-3}) t ({\rm s}) / 3\times10^{13}$,
where $t$ is time since the material had been shocked 
(Nozawa et al. 2006; Dwek et al. 1996).
This is proportional to the ionisation age 
$t_{\rm ion} =  n_e t$, assuming that $n_e \approx n_{\rm ion}$.
An ionisation age $t_{\rm ion} \approx 6\times10^{11}$ cm$^{-3}$ s 
was derived from soft X-ray spectra in 2018 (Ravi et al.
2021).
This ioniation age suggests that grain radii have been decreased 
by $\sim0.02$ $\mu$m with grains smaller than this being
completely destroyed in the shocked gas.

If dust grains are efficiently destroyed in the hot gas, a steady influx of dust into the shocked gas is needed to sustain the observed IR emission. 
This requires an increase in the volume of shocked gas, which can manifest itself by an increase in the X-ray emission measure (EM) defined as EM=$\int n_p n_e dV$, where $n_p$ is proton density and $V$ is the volume.
A two component model yields a $n_{\rm e}$ of $\sim$7500\,cm$^{-3}$ for the slow shock and 235\,cm$^{-3}$ for the fast shock in earlier days \citep[day 3829; ][]{Park.20046y}.
The EM of the soft component increased with time, more than expected by a simply expansion of the disk, until day$\sim$9000  \citep{Ravi.2021}.
After that, the EM of the soft component appeared to decrease, while that of the hard component continued to increase, with the rate changing at about day$\sim$9000.
There are two complications in interpreting the EM.
First, the shocks exited the ring on the east side around day$\sim$9000, so that the EM does not reflect a simple increasing volume anymore at this point.
Second,  \citet{Ravi.2021} mentions that a simple two component model analysis shows that the soft and hard components are becoming merged, and also the energy from the hard component contributes more to the total energy in post day$\sim$9000.
Because this day is coincident with the time when the dominant component shifts from the east to west side of the ring, it might be possible that fundamentally, the properties of shocks on the east and the west are different. 
It might be possible that the west side of the ring has a higher density, and causing harder shock energy, and also higher $n_{\rm e}$.
 However, with limited angular resolution in X-ray spectra, it is hard to evaluate.

\section{Conclusions}

Our MIR images of SN\,1987A show that the IR emission is 60\,\% dominated by dust emission from the west side of the circumstellar ring in recent years (2017--2022). 
That is in contrast to  IR images up to 2006, when the east side of the ring dominated the overall IR brightness (nearly 80\,\%).
The east side of the ring was hit by the blast waves earlier, causing them to brighten and led to the higher measured dust temperatures in the early 2000s.
After the shock waves passed the east side, the IR emission faded on that side.
Blast waves reached the west side of the ring later, and now the IR emission is dominated by the west side of the ring.
The shift of brightness from the east to the west is consistent with what was observed in the optical and the X-ray images.
The MIR observations also confirm that the IR/X-ray brightness ratio has been decreased in recent years.
This is mainly due to the increase of X-ray brightness, while the IR brightness has remained more or less constant within the uncertainties.

We suggest that the most likely cause for the decrease in the IR/X-ray ratio is that the  west side of the circumstellar ring has different properties from the east. 
It can be due to higher density, higher electron density or/and longer ionisation time scale.

 In summary, more precise models and more observations spanning a wider range of IR and X-ray wavelengths at higher resolution are required to quantitatively evaluate which effect(s), gas cooling, dust destruction, changes in gas density, or the exhaustion of new shocked gas and dust are responsible for the changes in the relative intensities of the global and local X-ray and IR emissions.

\section*{Acknowledgements}

We acknowledge  Dr K. Frank and D. Burrows for providing Chandra X-ray images of SN\,1987A, and Dr Kevin Volk for his advice on mid-infrared data analysis.
This work is based on observations collected at the European Southern Observatory under ESO programmes, 298.D-5023, 102.D-0245 and 106.D-2177.
Based on observations made with the NASA/DLR Stratospheric Observatory for Infrared Astronomy (SOFIA). SOFIA is jointly operated by the Universities Space Research Association, Inc. (USRA), under NASA contract NAS2-97001, and the Deutsches SOFIA Institut (DSI) under DLR contract 50 OK 0901 to the University of Stuttgart. 
Financial support for this work was provided by NASA through SOFIA 04-0016, 05-0050,  and 07-0064 issued by USRA.
%
Based on observations obtained at the international Gemini Observatory, a program of NSF's NOIRLab, which is managed by the Association of Universities for Research in Astronomy (AURA) under a cooperative agreement with the National Science Foundation, on behalf of the Gemini Observatory partnership: the National Science Foundation (United States), National Research Council (Canada), Agencia Nacional de Investigaci\'{o}n y Desarrollo (Chile), Ministerio de Ciencia, Tecnolog\'{i}a e Innovaci\'{o}n (Argentina), Minist\'{e}rio da Ci\^{e}ncia, Tecnologia, Inova\c{c}\~{o}es e Comunica\c{c}\~{o}es (Brazil), and Korea Astronomy and Space Science Institute (Republic of Korea).
This work is based [in part] on archival data obtained with the Spitzer Space Telescope, which is operated by the Jet Propulsion Laboratory, California Institute of Technology under a contract with NASA. Support for this work was provided by an award issued by JPL/Caltech.
Based on observations made with the NASA/ESA Hubble Space Telescope, and obtained from the Hubble Legacy Archive, which is a collaboration between the Space Telescope Science Institute (STScI/NASA), the Space Telescope European Coordinating Facility (ST-ECF/ESA) and the Canadian Astronomy Data Centre (CADC/NRC/CSA).
M.M.  acknowledges support from STFC Ernest Rutherford fellowship (ST/L003597/1) and STFC Consolidated grant (2422911).
Work by RGA was supported by NASA under award number 80GSFC21M0002.
M.J.B., and R.W. acknowledge support  from European Research Council (ERC) Advanced Grant SNDUST 694520, and 
HLG and P.C. acknowledge support from the European Research Council (ERC) in the form of Consolidator Grant {\sc CosmicDust} (ERC-2014-CoG-647939). 

\section*{Data avaiability}
The VISIR data are posted at ESO archive, and the VISIR and GEMINI reduced data will be available on request to the authors.



\bibliography{sn1987a_visir}
\bibliographystyle{mn2e}




\bsp	
\label{lastpage}
\end{document}